\documentclass[%
 reprint,
nofootinbib,
 amsmath,amssymb,
 aps,
]{revtex4-2}
\usepackage{graphicx}
\usepackage{dcolumn}
\usepackage{bm}

\usepackage{subcaption}
\usepackage[font=small,labelfont=bf,
   justification=Justified,
   format=plain]{caption} 
\usepackage{hyperref}
\hypersetup{colorlinks=true, citecolor=red, linkcolor=blue, urlcolor=blue}
\usepackage{amsmath}
\usepackage{bbm}
\usepackage{amsfonts}
\usepackage{amssymb}
\usepackage{latexsym}
\usepackage{graphicx}
\usepackage[english]{babel}
\usepackage{multirow}
\usepackage{float}
\usepackage{url}
\usepackage{slashed}
\usepackage{xcolor} 
\usepackage[utf8]{inputenc}
\usepackage{verbatim}
\usepackage{stackengine}
\usepackage{comment}

\def\sfrac#1#2{{\textstyle{#1\over #2}}}

\newcommand{\be}{\begin{equation}}
\newcommand{\ee}{\end{equation}}
\newcommand{\ba}{\begin{array}}
\newcommand{\ea}{\end{array}}
\newcommand{\bea}{\begin{eqnarray}}
\newcommand{\eea}{\end{eqnarray}}
\newcommand{\sss}{\scriptscriptstyle}
\newcommand{\nn}{\nonumber}

\usepackage{slashed}

\begin{document}
\rightline{CERN-TH-2024-137}

\preprint{APS/123-QED}

\title{Can Baby Universe Absorption Explain Dark Energy?}
\author{Varun Muralidharan}
\email{varun.muralidharan@mail.mcgill.ca}\thanks{ORCID: \href{https://orcid.org/0009-0004-8460-5230}{0009-0004-8460-5230}}
\author{James M. Cline}
\email{jcline@physics.mcgill.ca}
\thanks{ORCID: \href{https://orcid.org/0000-0001-7437-4193}{0000-0001-7437-4193}}
\affiliation{McGill University Department of Physics \& Trottier Space Institute, 3600 Rue University, Montr\'eal, QC, H3A 2T8, Canada}
\affiliation{CERN, Theoretical Physics Department, Geneva, Switzerland}

\begin{abstract}
It has been proposed that the accelerated expansion of the universe can be explained by the merging of our universe with baby universes, resulting in dark energy with a phantom-like equation of state.   However, the evidence in favor of it did not include the full set of cosmological observables.
Here we examine the implications of this model for both early and late universe cosmology using data from Planck collaboration, DESI 2024 and other experiments.  We find that the pure baby universe model gives a poor fit to current data.  Extending it to include a contribution from the cosmological constant, we find two allowed regions of parameter space: one close to $\Lambda$CDM, and another
with $\Lambda <0$ plus the exotic dark energy component.  The two regions can be significantly favored over $\Lambda$CDM, depending on the choice of supernova datasets, and they can ameliorate the Hubble tension to the level of $2\sigma$, depending on the supernova dataset.   The model with $\Lambda<0$ features an equation of state $w(a)$ with a pole singularity at early times. 

\end{abstract}

\maketitle


\section{Introduction}
Although the cosmological constant is the simplest description of dark energy, and still gives a good fit to cosmological data, there remains steady interest in possible deviations from time-independent dark energy, and many theories of modified gravity have been studied with respect to their ability to better describe the observations.  In particular, the Hubble tension problem has given impetus to certain classes of early dark energy, as have been reviewed in Ref.\ \cite{DiValentino:2021izs}.  These have the feature that the main modifications to the expansion history occur at high redshifts, and have been shown to be the most promising class of models for ameliorating the Hubble tension.

Many of the proposed modifications of gravity involve {\it ad hoc} functions of the Ricci scalar replacing the simple Einstein-Hilbert action for gravity, and are lacking any plausible ultraviolet completion.  In contrast, 
Refs.\ \cite{Ambjorn:2023hnt,Ambjorn:2024igl} (see also \cite{Ambjorn:2017cxu,Ambjorn:2023wxe}) put forward a model of quantum gravity that has a quite concrete and physical basis,
namely the absorption and emission of baby universes by the observed Universe.  It leads to a modified Friedmann equation for the expansion of the Universe, which depends only on a coupling constant $g$,  and whose sole energy contributions at late times is from the nonrelativistic matter (baryons plus dark matter); the accelerated expansion arises through a modification of the Friedmann equation rather than through an explicit dark energy contribution.

\begin{figure*}[t]
\begin{subfigure}{0.49\textwidth}
\includegraphics[width=0.8\linewidth, height=5cm]{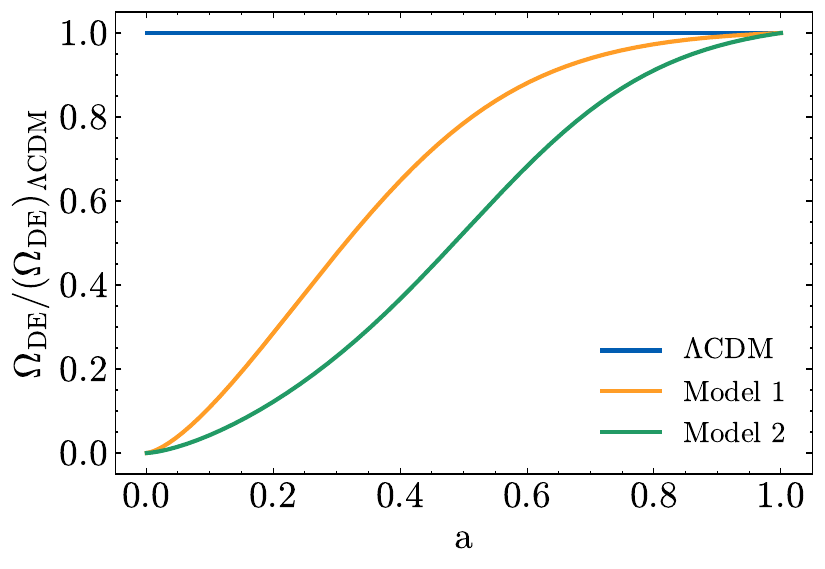} 
\label{fig:subim1}
\end{subfigure}
\begin{subfigure}{0.49\textwidth}
\includegraphics[width=0.8\linewidth, height=5cm]{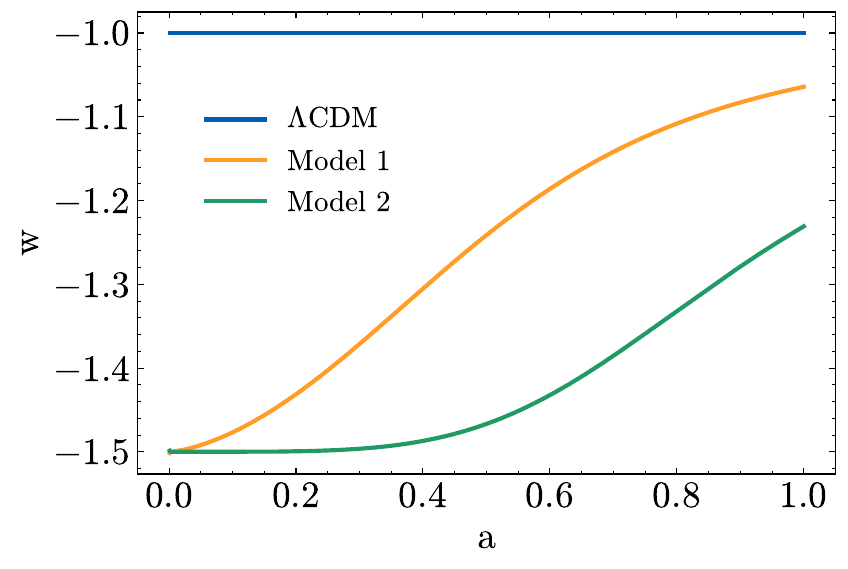}
\label{fig:subim2}
\end{subfigure}
\caption{Left: Dark energy density fraction $\Omega_{\rm DE}$ for the pure baby universe model ($\Lambda=0$) as a function of scale factor (a) relative to $\Lambda$CDM. Right: Equation of state of dark energy (w) as a function of scale factor. The curves were produced using Planck 2018 best fit values $\Omega_m h^2$ = 0.14072 and $H_0 = 67.32. $}
\label{fig:img1}
\end{figure*}

Ref.\ \cite{Ambjorn:2023hnt} found evidence in favor of this
model using a limited subset of available cosmological data, 
and concluded that it gives a good fit if the Hubble parameter agrees with late-time determinations, $H_0\cong 73$\,km/s/Mpc.
Their analysis was done without using CMB data, hence it is of interest to know how it fares against all available data.  We  modified the \texttt{Cobaya} \cite{Torrado:2020dgo} code in order to test the model's predictions with respect to
the standard set of observables, described in section \ref{constraints}.  In order to compare it to standard $\Lambda$CDM, we extend the model with 
a parameter $F_\Lambda$ that represents the fraction of the dark energy that comes from cosmological constant rather than from baby universe absorption.  We will show that $F_\Lambda$ must be greater than $0.64$ or less than $-0.16$ at 95\% confidence level;
hence the pure baby universe scenario with $F_\Lambda=0$ is ruled out.  However, we find some admixtures of cosmological constant and baby universe dark energy that can give significantly better fits than $\Lambda$CDM, depending on which supernova data are included.

\section{The Models}
We start with a recapitulation of the baby universe dark energy framework.  It is a minisuperspace model where the volume $v = a(t)^3/\kappa$  of a given topologically distinct universe (with $\kappa = 8\pi G)$ is
the dynamical degree of freedom.  Its 
conjugate momentum $p$ is related to 
$\dot a$ by
\be
    {\dot a\over a } = H = -\sfrac13 f'(p)\,.
    \label{pdef}
\ee
For a given 
function $f(p)$, Eq.\ (\ref{pdef}) can be solved to find $p$ as a function of $\dot a/a$.  The modified Friedmann equation is then given in terms of the energy density of matter and radiation,  omitting any vacuum energy, by 
\be
    f(p) = \kappa (\rho_{m} + \rho_r)\,.
    \label{feq}
\ee
The function $f(p)$ encodes an extra nonlinear term in the Hamiltonian of the
quantized minisuperspace model, that represents the effect of an initially disconnected region of space (``baby universe'') merging with, or being emitted by, the observable Universe at some time in the past, parametrized by the volume $v$ of the observed Universe.  One can picture it as a local interaction in which two universes merge into one, or one splits into two, with some interaction strength $g$, that has dimensions of [mass]$^3$.  The sign of the coupling determines whether the interaction represents merging ($g>0$) or splitting ($g<0$).\footnote{See footnote 5 of Ref.\ \cite{Ambjorn:2023hnt}}\   Its magnitude could naturally be exponentially suppressed
by the action of the gravitational instanton that describes the baby universe coupling
\cite{Giddings:1988wv}.
This could in turn explain the small magnitude of the dark energy, assuming that the cosmological constant vanishes.  We recall that Coleman suggested a mechanism by which baby universe interactions could select vanishing cosmological constant as the preferred state of the Universe \cite{Coleman:1988tj}.

This framework can reproduce the standard $\Lambda$CDM model if one turns off the coupling $g$ associated with the interaction of baby universes, and makes an appropriate choice of the function $f(p)$:
\begin{equation}
    f_\Lambda(p) \equiv \frac{3}{4}p^2 - \Lambda\,,
    \label{LCDM}
\end{equation}
which implies $p = -2H$.  The Friedmann equation (\ref{feq}) is then $H^2 = \kappa(\rho_m + \rho_r)/3 + \Lambda/3$, 
which is the expected result for vacuum energy 
\begin{equation}
    \rho_\Lambda = \Lambda/\kappa\, 
    \label{eq4}
\end{equation}
coming from a cosmological constant $\Lambda$.

Let us now consider 
a universe with vanishing cosmological constant $\Lambda = 0$, and merging of baby universes with a coupling constant $g>0$  as the sole source of dark energy. In this case, $f(p)$ is given by
\cite{Ambjorn:2023hnt}
\bea
     \label{mod1}
    f_1(p) &\equiv& -\frac{3}{4}(p+\alpha)\sqrt{(p-\alpha)^2 + 2\alpha^2}\,,\\
    \alpha &\equiv& g^{1/3}\,.\nn   
\eea
At early times, $p$ is large and 
one can expand Eq.\ (\ref{mod1}) in powers of $\alpha/p$, which leads to a simpler ansatz for $f(p)$,
\begin{equation}
    f_2(p) \equiv \frac{3}{4}\left(p^2 + \frac{2g}{p}\right)\,,
    \label{mod2}
\end{equation}
that was also considered by Refs.\ \cite{Ambjorn:2023hnt,Ambjorn:2024igl}.
By comparing to Eq.\ (\ref{LCDM}), one sees that
the baby universe model resembles $\Lambda$CDM,
but with dark energy that is 
growing with time, since $p\sim -H$ is decreasing.
For convenience, we will refer to the two alternatives (\ref{mod1}) and (\ref{mod2}) as Model 1 and Model 2, respectively.  Although Model 2 will turn out to give a worse fit to the data, it provides a simple qualitative understanding of the effects of the exotic dark energy source.

The above formalism can alternatively be understood in terms of the 
ordinary Friedmann equation $H^2 = \kappa\rho_{\rm crit}/3$, by defining an equivalent (formal) dark energy density $\rho_{f} = \rho_{\rm crit} - (\rho_m+\rho_r)$, with
$\rho_{\rm crit} = 3H^2/\kappa = f'(p)^2/(3\kappa)$, from Eq.\ (\ref{pdef}),
and $(\rho_m+\rho_r) = f(p)/\kappa$, from Eq.\ (\ref{feq}).
This gives
\be
    \rho_{f} = {1\over\kappa}\left(\sfrac{1}{3}f'(p)^2 - f(p)\right)
    \label{eq7}
\ee
where $\rho_f$ is the dark energy density coming from baby universe absorption. This form is useful for comparing the baby universe model to other dark energy models,
including the cosmological constant.
The value of the coupling constant $g$ (alternatively $\alpha$), and the present value $p_0$ of $p$, are related to the observables $\rho_{m0}$
and $H_0$ (the current matter density and Hubble parameter) through
\bea
      f(p_0) &=& \kappa\rho_m(v_0) = \kappa\rho_{m0}\nn \\
      f'(p_0) &=& -3H_0\,.
      \label{initcond}    
\eea
 Although it has a negligible effect, we also included the contribution from the radiation density in $\rho_{m0}$. 

Equations (\ref{initcond}) can be numerically solved to determine $g$ and $p_0$.  Therefore  Models 1 and 2 do not have any additional physics parameters beyond those already present in $\Lambda$CDM: $\Lambda$ is replaced by $g$.  For greater generality, we define Model 3 by introducing a parameter
$F_\Lambda$ that interpolates between Model 1, which is supposed to be the more accurate representation of baby universe absorption, and $\Lambda$CDM, using the modified Friedmann equation
\be
    f_1(p) = \kappa(\rho_m + F_{\Lambda}\rho_{\rm \sss DE})\,. \\
    \label{eq9}
\ee
Here $\rho_{\rm \sss DE}$ is the total dark energy density in the universe ($\rho_{\rm crit} - \rho_m$). This is equivalent to defining $f_3(p) \equiv f_1(p) - F_{\Lambda}\Lambda$
and writing the modified Friedmann equation as $f_3(p) = \kappa(\rho_m+\rho_r)$, as in Eq.\ (\ref{feq}).
When $F_{\Lambda} = 0$ , $f_3(p)$ reduces to $f_1(p)$ and all of the dark energy density comes from absorption  of baby universes. On the other hand, when $F_{\Lambda} = 1$, $f_1(p) = \kappa\rho_{\rm crit}$ and so from Eqs.\ (\ref{eq7}) and (\ref{pdef}), $\kappa\rho_{\rm\sss DE} = 3H^2-\kappa\rho_{\rm crit} + \Lambda = \Lambda$ where $\rho_{\rm crit} = 3H^2/\kappa$ is the critical density of the universe. This is the same as Eq.\ (\ref{eq4}),
which implies that all of the dark energy density is in the form of cosmological constant.   Moreover, it is possible to have 
$F_\Lambda >1$, counteracted by $g<0$ (baby universe emission).  With this extra parameter, we will be able to derive limits on the fraction of dark energy that can be attributed to baby universe absorption, or preferred models that have the combined two sources of dark energy.

\section{Analysis}

The dark energy density of the baby universe models is quite different from $\Lambda$CDM at high redshifts, altering the expansion history of the universe. 
As shown in Fig.\ \ref{fig:img1} (left), 
there is less dark energy at high $z$, which can enhance the growth of structure.
Equivalently, the equations of state of the two dark energy models are more negative than $-1$, as we illustrate in Fig.\ \ref{fig:img1} (right). The modified background evolution was implemented in the Boltzmann code \texttt{CAMB} \cite{Lewis:1999bs}. We do not consider any perturbations in the dark energy density that might arise during the absorption of baby universes, since the description of such processes has not been worked out.

Since the baby universe models provide primarily late dark energy, one can
expect Models 1-3 to be in good agreement with the CMB power spectrum  and temperature auto-correlation spectrum.  This
will be borne out in the following analysis.  Instead, baryon acoustic oscillations (BAO) and supernovae turn out to drive the
constraints.

\subsection{Universe with $\Lambda = 0$}
\label{constraints}
To incorporate up-to-date constraints on the cosmological
observables, we used the publicly available MCMC sampler \texttt{Cobaya} \cite{Torrado:2020dgo}. Convergence of the chains was monitored via the Gelman-Rubin statistic \cite{Gelman:1992zz}, demanding $R-1 \lesssim 0.02$.  The likelihoods included in our analysis are:
\begin{itemize}
    \item \textbf{Planck:} Measurements of CMB temperature and polarizations anisotropies and cross-correlations from Planck 2018 \cite{Planck:2018vyg};
    \item \textbf{Lensing:} CMB lensing potential power spectrum, reconstructed from the CMB temperature four-point function from Planck 2018 \cite{Planck:2018lbu};
    \item \textbf{BAO:} Distance measurements from baryon acoustic oscillations. We follow the methodology described in the DESI 2024 paper combining data from DESI and SDSS. 
    \begin{itemize}
        \item At $z<0.6$, we use SDSS results from Main Galaxy Survey ($z_{\rm eff} = 0.15$) and DR12 BAO ($z_{\text{eff}}=0.38, 0.51$) since SDSS has a larger effective volume \cite{Ross:2014qpa, BOSS:2016wmc},
        \item At $z>0.6$, we use results from DESI 2024 LRG, LRG+ELG, ELSGs, QSO since DESI has a larger effective volume \cite{DESI:2024lzq, DESI:2024mwx, DESI:2024uvr},
        \item For Ly$\alpha$ ($z_{\rm eff} = 2.33$), we use the combined result from SDSS + DESI 2024; 
    \end{itemize}

    \item  \textbf{DES}: Cosmic shear and clustering measurements from Dark Energy Survey  year 1 results \cite{DES:2017myr, DES:2017tss, DES:2017qwj}; 
    \item  \textbf{SN:} Distance measurements of type Ia supernovae from Pantheon+ consisting of 1701 distinct lightcurves of 1550 supernovas ranging in redshift from $z=0.001$ to 2.26 \cite{Brout:2022vxf}. For the second part of our analysis we also consider Union3 compilation of 2087 SN Ia \cite{Rubin:2023ovl}  and a sample of 1635 photometrically-classified SN $(0.1<z<0.3)$ from Dark Energy Survey year 5 data release \cite{DES:2024tys}. 
\end{itemize}

\begin{center}
\begin{table}

\begin{tabular}{|c|c|c|c|}
\hline
Parameter & $\Lambda$CDM & Model 1 & Model 2 \\
 \hline \hline
$\Omega_{b}h^2$ &0.0227 & 0.0224&0.0221\\
$\Omega_{c}h^2$ &0.1171 & 0.1198&0.1231\\
$100 \theta_{\rm MC}$ & 1.0413& 1.0409& 1.0405\\
$\tau$ &0.0593&0.0527&.0439\\
$\ln(10^{10}A_s)$ & 3.0493 & 3.0409&3.0268\\
$n_s$ &0.9720 &0.9663& 0.9569\\

$H_0                       $ & 68.72 & 71.33& 74.96 \\
\hline
$\chi^2_{\rm lensing}      $ & 9.6& 8.7& 11.7 \\

$\chi^2_{\rm SN}          $ & 1482.9& 1485.5& 1566.4 \\

$\chi^2_{\rm BAO}      $ & 10.4& 15.2& 27.5 \\

$\chi^2_{\rm DES}          $ & 507.6 & 508.7 & 512.9 \\

$\chi^2_{\rm CMB}          $ & 2769.9 & 2765.4& 2775.0\\
 \hline
$\rm Total ~\chi^2             $ & 4780.4& 4783.5 & 4893.5\\
$\Delta \chi^2             $ & 0& 3.1 & 113.1\\
 \hline  
\end{tabular}
\caption{Best-fit cosmological parameters for $\Lambda$CDM and the two baby universe models, Model 1 and Model 2, found in the MCMC scan. The total $\chi^2$ for each dataset is given in the last few rows and $\Delta \chi^2 = \chi^2 - \chi^2_{\Lambda\rm CDM}$. } 
\label{table1}
\end{table}
\end{center}

We computed the $\chi^2$ values for the first two models, which have no free parameters, and compared them with $\Lambda$CDM. The results are shown in Table \ref{table1}. $\Lambda$CDM is mildly preferred over Model 1, while observations strongly exclude Model 2.  Even though Model 1 provides modest improvement over $\Lambda$CDM for CMB observations, it is in tension with BAO and SN data, which probe the late-time acceleration. 
We repeated the analysis using only SDSS BAO data \cite{BOSS:2016wmc, Ross:2014qpa} instead of DESI 2024 BAO but obtained similar results, as can be seen in  Table \ref{table2}.  
The models become further strongly disfavored with respect to $\Lambda$CDM by choosing different 
supernova datasets,  Union3 and DESY5, also shown in Table \ref{table2}. 

\begin{figure}[t]
\!\!\!\!\  \includegraphics[height=4.6cm]{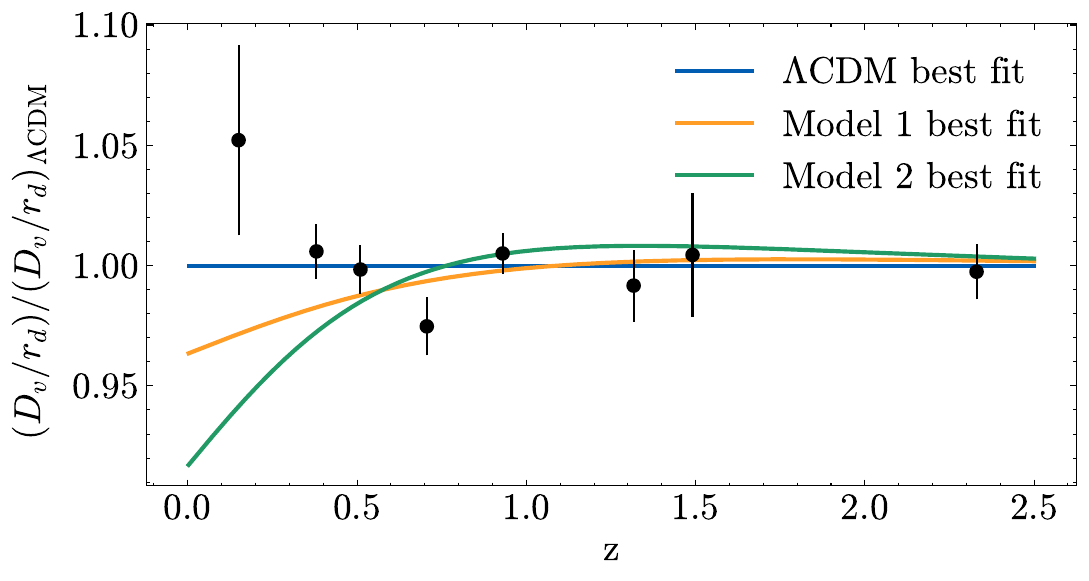}
    \caption{Measurements of BAO distance scales versus redshift, parametrized as the ratio of the angle-averaged distance $D_v(z) = (zD_{\rm M}^2D_{\rm H})^{1/3}$ to the sound
horizon at the baryon drag epoch $r_{\rm d}$. The blue, yellow and green lines respectively show the prediction for $\Lambda$CDM, Model 1 and Model 2 best-fit values obtained in Table \ref{table1} relative to the $\Lambda$CDM prediction.}
    \label{BAOfig}
\end{figure}

Perhaps unsurprisingly, Model 2, which is an early-time approximation for the more exact Model 1, fails badly in reproducing late-time observations.  But even Model 1 
is in conflict with BAO observations at redshifts $z\lesssim 0.5$, as shown in Fig.\ \ref{BAOfig}.  Comparing to Fig.\ 3 of Ref.\ \cite{Ambjorn:2023hnt} (based on data listed in Table 1 of Ref.\ \cite{Ambjorn:2022wro}), which found a good fit to BAO data,
we see that the adopted data look quite different at low $z$
between that analysis and the present one.  The data responsible
for the discrepancy are those at low redshifts $z\lesssim 0.5$.
Ref.\ \cite{Ambjorn:2023hnt} cited surveys dating from 2011 and earlier, whereas we are using Refs.\ \cite{Ross:2014qpa, BOSS:2016wmc}, from 2014-2016, which are provided with 
\texttt{Cobaya}.    Hence our discrepant conclusions relative to Ref.\ \cite{Ambjorn:2023hnt} can be attributed to the fact that the data have evolved since 2011.

\begin{center}
\begin{table}

\begin{tabular}{|c|c|c|c|}
\hline
Survey & $\Lambda$CDM & Model 1 & $\Delta \chi^2$\\
 \hline

PantheonPlus & 4780.4 & 4783.5 & 3.1\\

Union3 & 3323.0 & 3340.2 & 17.2 \\

DESY 5 & 4944.0 & 4978.9 & 34.9\\
\hline
SDSS BAO & 4775.5 & 4780.3 & 5.2 \\
\hline
No CMB & 1968.1 & 1981.5 & 13.4 \\
 \hline  
\end{tabular}
\caption{Total $\chi^2$ values determined by reweighting the chains using supernova data from different surveys, and $\Delta \chi^2 = \chi^2 - \chi^2_{\Lambda\rm CDM}$. The first three rows include the same data as in Table \ref{table1} but the supernova data from Pantheon+ are replaced by those of Union3 or DESY 5, respectively. The penultimate row includes datasets from Table \ref{table1} but the BAO data from DESI are replaced by those from SDSS. The last row shows the total $\chi^2$ determined by excluding CMB data from Planck 2018. } 
\label{table2}
\end{table}
\end{center}

\subsection{Universe with $\Lambda >0$}
\begin{figure}
    \centering
    \includegraphics[height=8cm]{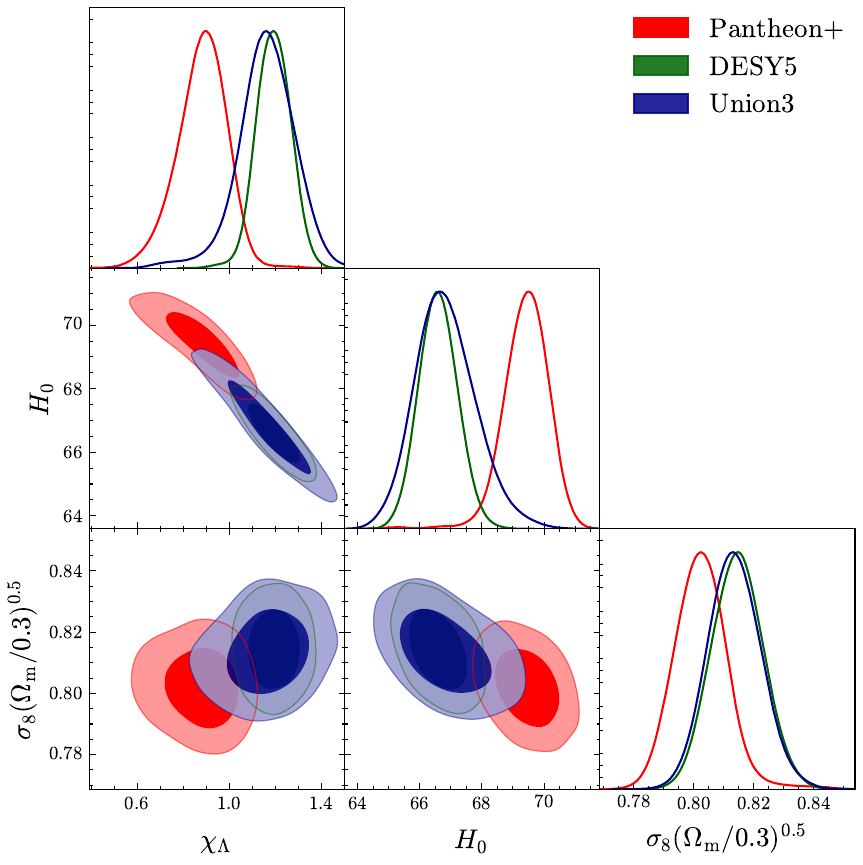}
    \caption{Marginalized posterior on $F_{\Lambda}$, $H_0$ and $\sigma_8$ obtained in the full MCMC scan for Model 3 with $\Lambda>0$. The dataset used here is Planck + Lensing + DESI BAO + DES with the supernova dataset mentioned.}
    \label{fig3}
\end{figure}

To study the landscape of theories for a universe with both a positive cosmological constant $\Lambda >0$ and baby universe
coupling $g>0$, we use Model 3 defined in Eq.\ (\ref{eq9}).
$\Lambda$CDM and Model 1 are both contained in this description and occur when $F_{\Lambda} = 1$ and $F_{\Lambda}=0$ respectively. We follow the procedure outlined earlier to determine constraints on $F_{\Lambda}$ using \texttt{Cobaya.} 
The
1-$\sigma$ and 2-$\sigma$ allowed regions are shown in Fig.\ \ref{fig3}, for the three choices of SN datasets,  and the $95\%$ C.L.\ results are given in Table \ref{table3}.
In addition to $F_\Lambda$, we show distributions of $H_0$ and the clustering 
parameter $S_8 = \sigma_8(\Omega_m/0.3)^{0.5}$, which has also exhibited a mild tension between early- and late-time observables \cite{DiValentino:2020vvd}. The estimated $\Lambda$CDM value ($S_8 = 0.834 \pm 0.016$) from Planck \cite{Planck:2018lbu} is in a 3$\sigma$ tension with the KiDS-1000 reported value ($S_8 = 0.766^{+0.020}_{
-0.014})$ \cite{Heymans:2020gsg}. The best-fit DESY5 curve ($S_8 = 0.815 \pm 0.016)$ reduces the tension to the 2$\sigma$ level.

\begin{center}
\begin{table}

\begin{tabular}{|c|c|c|c|}
\hline
Parameter & Pantheon+ & DESY5 & Union3 \\
 \hline
 \hline
$H_0                       $ & 69.45 & 66.60 & 66.71\\
$F_{\Lambda}$ & $0.86$ &  $1.18$ & $1.17$ \\
\hline
$\chi^2_{\rm lensing}      $ & 9.2& 9.9 & 9.5\\

$\chi^2_{\rm SN}          $ & 1483.5& 1638.7 & 23.2\\

$\chi^2_{\rm BAO}      $ & 10.8& 11.3 & 11.3\\

$\chi^2_{\rm DES}          $ & 508.1& 509.1 & 509.3\\

$\chi^2_{\rm CMB}          $ & 2767.5& 2768.2 & 2768.1\\
 \hline
$\Delta \chi^2             $ & $-1.3$& $-6.8$ & $-1.6$\\
 \hline  
\end{tabular}
\caption{Best-fit cosmological parameters for baby universe Model 3 including a cosmological constant $\Lambda>0$. The full dataset includes Planck + Lensing + DESI BAO + DES along with the supernova dataset indicated in the table.} 
\label{table3}
\end{table}
\end{center}

Using the Pantheon+ supernova data, the results indicate that most of the dark energy today is in the form of a cosmological constant, with only a small fraction of dark energy, if any, coming from the merging of baby universes. The pure baby universe scenario $F_{\Lambda} = 0$ (Model 1) is disfavored by more than 2$\sigma$.  

\begin{figure}
    \centering
    \includegraphics[height=5cm]{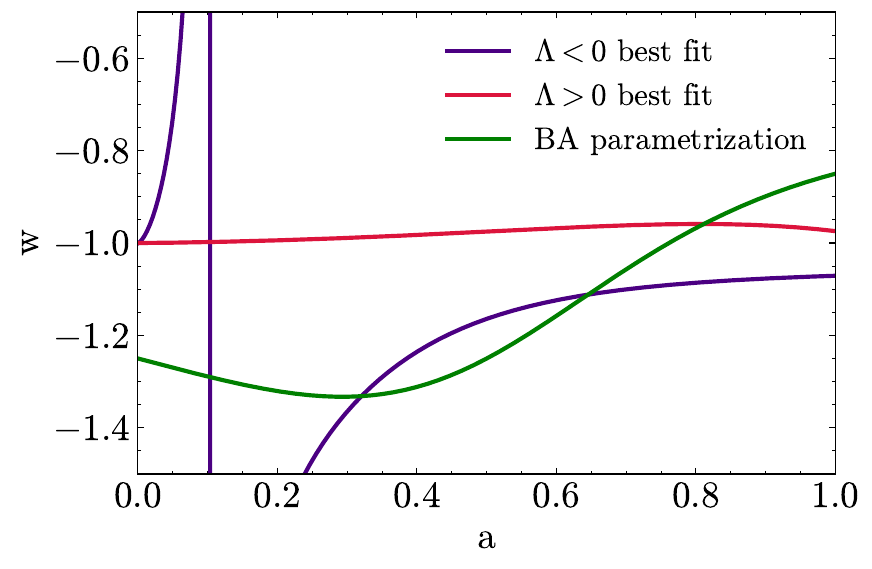}
    \caption{Equation of state (w) of dark energy as a function of scale factor (a) for Model 3. $\Lambda<0$ best-fit is shown in purple and $\Lambda>0$ best-fit is shown in crimson. Barboza-Alcaniz parametrization for best-fit values from Ref.\ \cite{Giare:2024gpk} ($w_0 = -0.84, w_a = -0.53)$ is shown in green.}
    \label{fig4}
\end{figure}

The best-fit value from DESY5 indicates that $F_{\Lambda} \approx 1.2> 1$, which implies that the baby universe coupling constant $g<0$, to maintain spatial flatness.  This corresponds to
a universe with excess cosmological constant,
that is compensated by the emission of baby
universes.  The full dataset + DESY5 offers a significant improvement over $\Lambda$CDM with $\Delta\chi^2 = -6.8$, as shown in Table \ref{table3}.
The equation of state of Model 3 with $\Lambda>0$ is  close to $-1$, as shown in Figure \ref{fig4}, because most of the dark energy is coming from the cosmological constant, with only a small contribution from the exotic dark energy source.

\subsection{Universe with $\Lambda<0$}
\begin{figure}
    \centering
    \includegraphics[height=4.5cm]{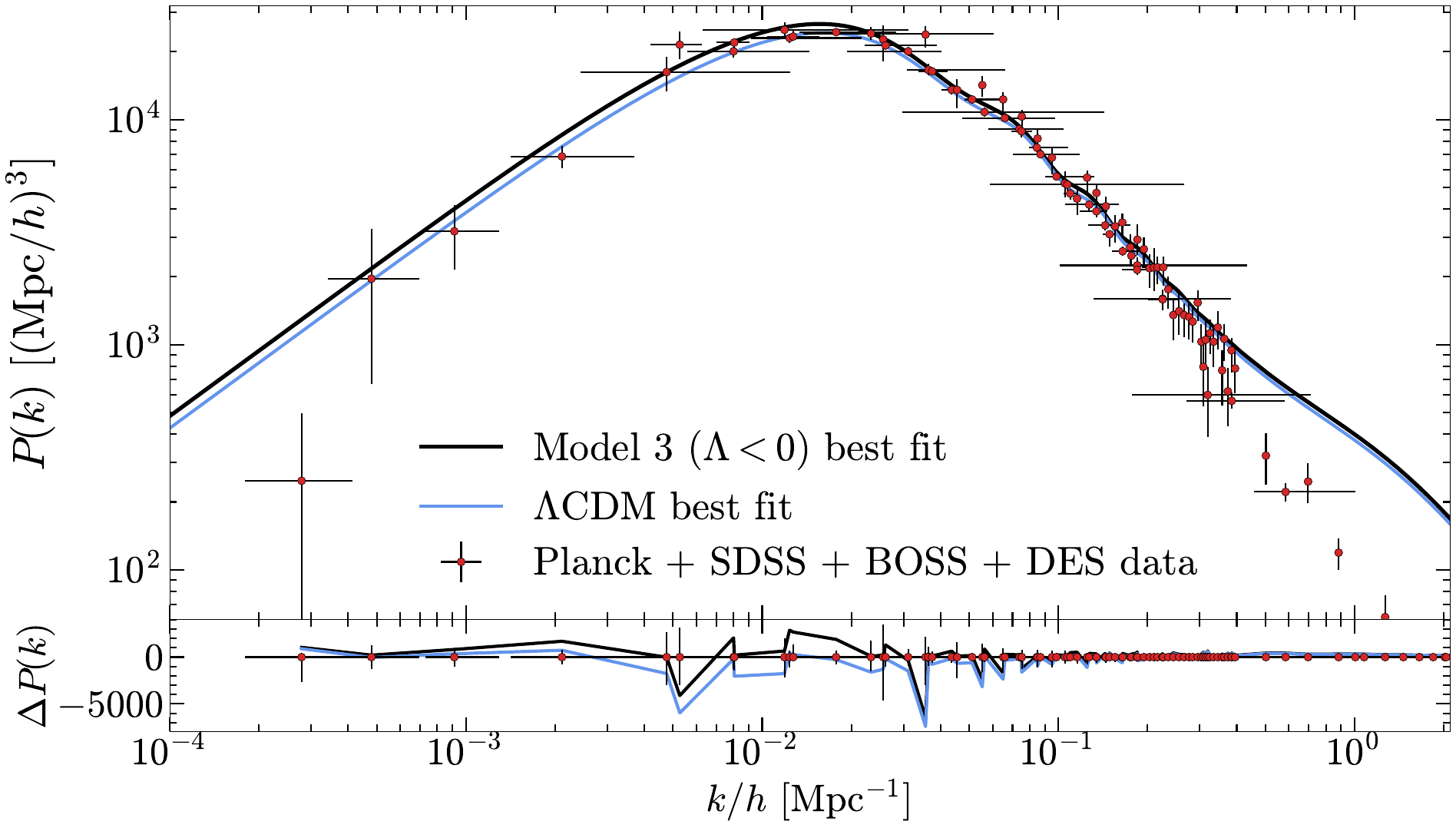}
    \caption{CMB matter power spectrum predictions for Model 3 and $\Lambda$CDM best fit values.}
    \label{cmb}
\end{figure}

Next we consider Model 3 in the region of parameter space where the cosmological constant is negative. Following the same procedure as above, we find that this can provide a significantly better fit than $\Lambda$CDM  when using the Pantheon+ dataset,  as shown in Table \ref{table4}.
In Figure \ref{cmb} we plot the CMB matter power spectrum predictions for Model 3 and the $\Lambda$CDM best fit values.  From the residuals, one sees that the improvement in $\chi^2$ is due to intermediate scales $k/h \sim 10^{-3}-10^{-1}$\,Mpc$^{-1}$.  

We note that the dark energy equation of state of $w(a)$  for Model 3 with negative $\Lambda$ has unusual behavior, diverging at $a\sim 0.1$ before entering the phantom-like regime with $w<-1$. This is due to the dark energy contribution changing sign, while its pressure remains negative.  In Figure \ref{fig4}, we compare $w(a)$  of the two scenarios to the Barboza-Alcaniz (BA) parametrization which has the form
\begin{equation}
    w(a) = w_0 + w_a\times \frac{1-a}{a^2 + (1-a)^2}.
\end{equation}
This parameterization was shown in Ref.\ \cite{Giare:2024gpk} to give an optimal fit to data with respect to the recent DESI BAO indications of evolving dark energy \cite{DESI:2024kob}.

The preference for Model 3, however, is not robust against choosing a different supernova dataset such as DESY5 or Union3, as demonstrated in Table \ref{table4}. Pantheon+ and Union3 have roughly 1360 supernova in common but differ in their analysis methodology. DESY5 has 194 supernova in common with the two but has an additional 1500 photometrically-classified SN Ia at higher redshifts.  In the concluding section we comment on a possible explanation
for the discrepancy between the DESY5 and other SN datasets. 

\begin{figure}
    \centering
    \includegraphics[height=6cm]{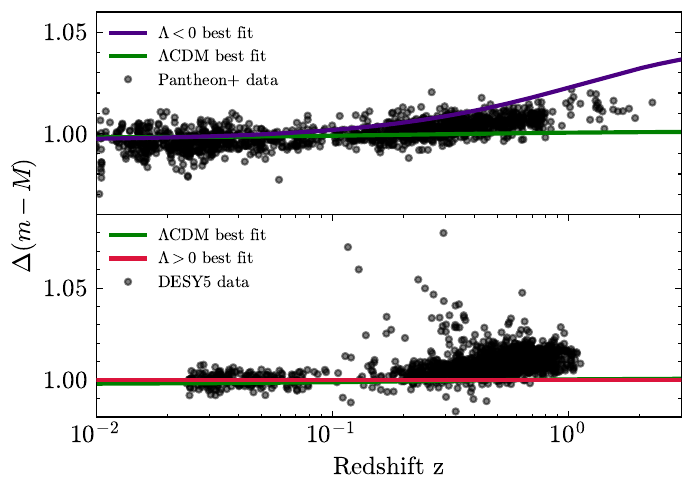}
    \caption{Distance modulus predictions for $\Lambda$CDM (green curves), Model 3 ($\Lambda<0$, top and $\Lambda>0$, bottom) best-fit. Here $\Delta(m-M) = (m-M)/(m-M)_{\Lambda>0}$ and $m-M = 5\log_{10}d_L + 25$ where $d_L$ is the luminosity distance. Pantheon+ and DESY5 $u$-band magnitude measurements for Type 1A SN are denoted by the black dots.}
    \label{dl}
\end{figure}

In Figure \ref{dl} we compare the distance modulus predictions for Model 3 for the two cases against $\Lambda$CDM. One sees that the improvement in $\chi^2$ for $\Lambda<0$ comes from intermediate redshifts  $z \sim 0.1-3.0$.  For $\Lambda>0$, the slight improvement in the fit is due to the outliers in the data. 

\begin{figure}[t]
    \centering
    \includegraphics[height=8cm]{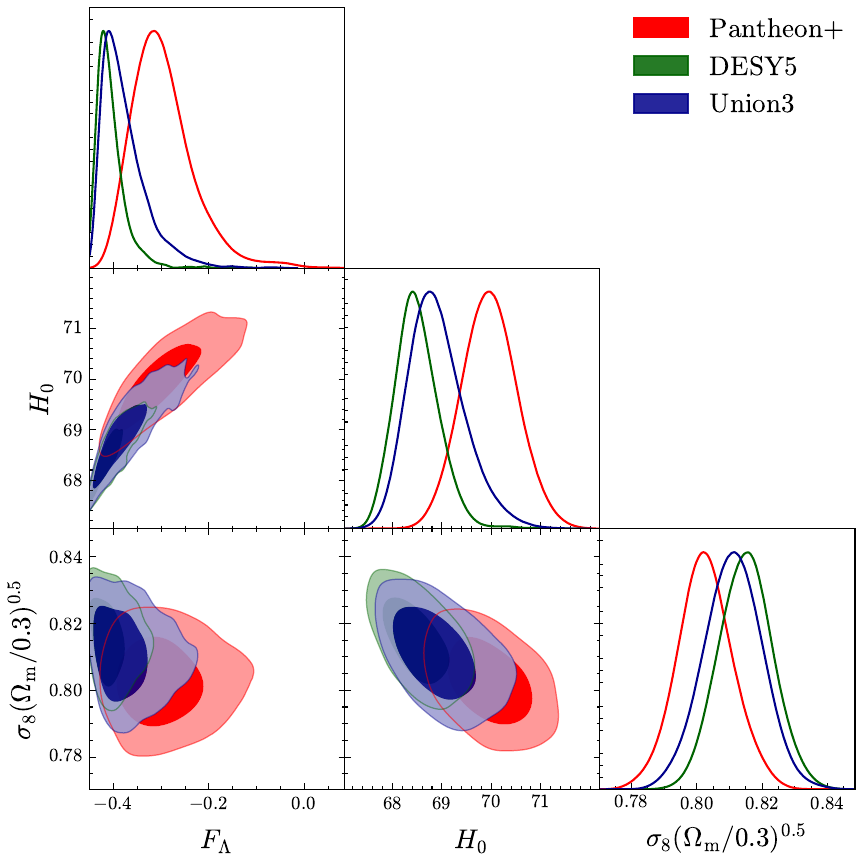}
    \caption{Marginalized posterior on $F_{\Lambda}$, $H_0$ and $\sigma_8$ obtained in the full MCMC scan for Model 3 with $\Lambda<0$.}
    \label{fig5}
\end{figure}

Figure \ref{fig5} shows the marginalized posterior on $F_{\Lambda}, H_0 ~\text{and} ~S_{8}$ for the different supernova dataset choices. It demonstrates the preference for $-0.5\lesssim F_\Lambda\lesssim -0.1$, and values of $H_0$ closer to the SN-preferred value in the case of Pantheon+ data.  Values of  $S_8\lesssim 0.82$ are preferred in this case.

\begin{center}
\begin{table}

\begin{tabular}{|c|c|c|c|}
\hline
Parameter & Pantheon+ & DESY5 & Union3 \\
 \hline
 \hline
$H_0                       $ & 69.90 & 68.15 & 68.54\\
$F_{\Lambda}$ & $-0.31$ &  $-0.43$ & $-0.41$ \\
\hline
$\chi^2_{\rm lensing}      $ & 8.7& 8.9 & 8.8\\

$\chi^2_{\rm SN}          $ & 1478.3& 1648.6 & 28.8\\

$\chi^2_{\rm BAO}      $ & 11.9& 11.0 & 10.4\\

$\chi^2_{\rm DES}          $ & 507.9& 509.9 & 509.3\\

$\chi^2_{\rm CMB}          $ & 2765.1& 2765.3 & 2765.6\\
 \hline
$\Delta \chi^2             $ & $-8.5$& $-0.3$ & $-0.1$\\
 \hline  
\end{tabular}
\caption{Best-fit cosmological parameters for baby universe model 3, including a cosmological constant $\Lambda<0$. The full dataset includes Planck + Lensing + DESI BAO + DES along with the supernova data indicated in the table.} 
\label{table4}
\end{table}
\end{center}

The value of the Hubble parameter obtained in the full MCMC scan for Model 3 with negative $\Lambda$ is 69.9 $\pm$ 1.1 $\text{km s}^{-1}\text{Mpc}^{-1}$ with 95\% CI. The latest distance ladder measurement of $H_0$ by the SH0ES collaboration is 73.2 $\pm$ 1.3 $\text{km s}^{-1}\text{Mpc}^{-1}$ at 65\% CI. Thus, Model 3 alleviates the Hubble tension to within $2\sigma$ of the local measurement. The best-fit $S_8 = 0.803 \pm 0.018$ also eases the tension to within $1\sigma$ of the KiDS-1000 reported value. 

\section{Discussion}
In this paper, we have systematically scrutinized the baby universe absorption model for accelerated expansion of the universe,  as proposed in Refs. \cite{Ambjorn:2023hnt, Ambjorn:2024igl}. Using data from various experiments, we were unable to find a preference for either of the two pure baby universe models over $\Lambda$CDM as $\Delta \chi^2 = 3.1$ and 113.1 respectively. We also reevaluated the chains using different supernova and BAO datasets but all of them reiterated the same fact.

Allowing for the cosmological constant to be different from 0 using a new parameter $F_{\Lambda}$ that we define as the fraction of the dark energy density in a cosmological constant $\Lambda$, we find two allowed regions of parameter space: one close to $\Lambda$CDM (best-fit $F_{\Lambda} = 0.86$) and another with negative $\Lambda$ (best-fit $F_{\Lambda} = -0.31$). Both of the regions are exclude the point $F_{\Lambda} = 0$ with 95\% CI which corresponds to the pure baby universe model that was initially proposed. The former has a $\Delta \chi^2 = -6.8$ using DESY5 supernova data while the latter one has a $\Delta \chi^2 = -8.5$ using Pantheon+ data. Both models are therefore significantly preferred over $\Lambda$CDM. The latter one predicts a higher value of $H_0 = 69.90$ and thereby ameliorates the Hubble tension to within 2$\sigma$ of the local measurement. 

Recently it was suggested that the DESY5 supernova data may suffer from a systematic 0.04 magnitude error in going between low and high redshifts, which is not observed in the Pantheon+ data, and correcting for this removes the evidence from DESI BAO masurements for evolving dark energy \cite{Efstathiou:2024xcq}.
On this basis, one might doubt the preferred region we found with $\Lambda>0$, which is driven by the DESY5 data.  However, our other preferred region with $\Lambda <0$ is instead driven by the Pantheon+ data, and does not suffer from this caveat.  

Although the $\Lambda<0$ baby universe model is not economical, needing two separate sources of dark energy, it makes the interesting point that an unusual equation of state dependence $w(a)\sim c/(a-a_0)+\dots$, having a pole at some (early) value of $a$, can give a good fit to present data.  The divergence is a consequence of the dark energy density changing sign at some redshift,  even though the total energy density remains positive.  Such behavior is not captured by standard parametrizations of $w(a)$ such as those considered in Ref.\ \cite{Giare:2024gpk}.

\medskip
\textbf{Acknowledgments.} We thank Jan Ambjorn, Robert Brandenberger, Kevin Langhoff, and Jon Sievers  for helpful discussions. We are grateful to the CERN Theory department for its kind hospitality while this work was being completed.  We thank Juan Gallego for assistance with the computing cluster. JC was supported by NSERC
(Natural Sciences and Engineering Research Council,
Canada). VM was partially supported through a MITACS Graduate Fellowship.

\bibliography{references}
\bibliographystyle{utphys}

\end{document}